\documentclass{desyproc}
\begin{document}
\title{Aspects of Higgs Production at the LHC}

\def\beq{\begin{equation}}
\def\eeq{\end{equation}}
\def\bea{\begin{eqnarray}}
\def\eea{\end{eqnarray}}
\newcommand{\bas}{\bar{\alpha}_S}
\newcommand{\as}{\alpha_S}
\newcommand{\bra}[1]{\langle #1 |}
\newcommand{\ket}[1]{|#1\rangle}
\newcommand{\bracket}[2]{\langle #1|#2\rangle}
\newcommand{\intp}[1]{\int \frac{d^4 #1}{(2\pi)^4}}
\newcommand{\mn}{{\mu\nu}}
\newcommand{\tr}{{\rm tr}}
\newcommand{\SP}{\langle \mid S^2 \mid\rangle}
\newcommand{\Tr}{{\rm Tr}}
\newcommand{\T} {\mbox{T}}
\newcommand{\braket}[2]{\langle #1|#2\rangle}
\newcommand{\Lb}{\left(}
\newcommand{\Rb}{\right)}
\newcommand{\m}{\margi }
\newcommand{\lash}[1]{\not\! #1 \,}
\newcommand{\nn}{\nonumber}
\newcommand{\D}{\partial}
\newcommand{\h}{\frac{1}{2}}
\newcommand{\g}{{\rm g}}
\newcommand{\x}{\vec{x}}
\newcommand{\z}{\vec{z}}
\newcommand{\kv}{\vec{k}}
\newcommand{\qv}{\vec{q}}
\newcommand{\lv}{\vec{l}}
\newcommand{\mv}{\vec{m}}
\newcommand{\y}{{\cal Y}}
\newcommand{\el}{{\cal L}}
\newcommand{\A}{{\cal A}}
\newcommand{\Ka}{{\cal K}}
\newcommand{\al}{\alpha}
\newcommand{\be}{\beta}
\newcommand{\ep}{\varepsilon}
\newcommand{\ga}{\gamma}
\newcommand{\de}{\delta}
\newcommand{\De}{\Delta}
\newcommand{\et}{\eta}
\newcommand{\ka}{\vec{\kappa}}
\newcommand{\la}{\lambda}
\newcommand{\ph}{\varphi}
\newcommand{\si}{\sigma}
\newcommand{\ro}{\varrho}
\newcommand{\Ga}{\Gamma}
\newcommand{\om}{\omega}
\newcommand{\La}{\Lambda}
\newcommand{\tG}{\tilde{G}}
\def\pom{{I\!\!P}}
\def\reg{{I\!\!R}}


\author{{\slshape Errol Gotsman}\thanks{e-mail: gotsman@post.tau.ac.il}
\\[1ex]
Department of Particles Physics, School of Physics and Astronomy\\
Raymond and Beverly Sackler Faculty of Exact Science\\
Tel Aviv University, Tel Aviv 69978, Israel}

\contribID{gotsman_errol}

\maketitle 

\begin{abstract}
We discuss the main features and predictions of the GLMM model, which is 
based on a QCD motivated theoretical approach, and successfully describes 
the experimental data on total, elastic and diffractive cross sections. In
 addition we
 calculate the survival probability for a SM Higgs at the LHC, and compare 
our 
results with those of the Durham group.
\end{abstract}

\section{Introduction}
    Over the past few years the subject of "soft physics" has reemerged 
from the shadows, and has aroused the interest of the phenomenological 
community. This in no small way, due to the realization that the 
calculation of the probability of detecting a diffractive hard pQCD 
process e.g. Higgs 
production at the LHC, also depends on the underlying secondaries which 
are produced by "soft" rescattering. Central diffractive production e.g.
(Higgs boson, 2 jets, 2 $\gamma$'s, $\chi_{c}$), are accompanied by  
gaps 
in rapidity, between the two outgoing projectiles, and the centrally 
produced 
particles, which makes their detection easier. The subject of the survival
of these rapidity gaps was initiated over twenty years ago \cite{SP1}, 
and has been refined over the interim period \cite{GLMM}, \cite{KMR09}. 

\section{Details of our Model}
 The details on which our two channel (GLM) model is based i.e. the 
Good-Walker (G-W)
mechanism \cite{GW}  can be found in \cite{GLMM}. See also U. Maor's talk 
in these proceeedings. It is well known that G-W neglects the diffractive 
production of large mass states (Mueller diagrams \cite{Muell1}), and 
  to successfully describe the diffractive data, these need to be 
included by adding the relevant triple 
Pomeron contributions (see L.H. diagrams in Fig.1).
 In addition it is also necessary to include
diagrams containing Pomeron loops, as shown in the R.H. diagrams in Fig.1.
\begin{figure}
\begin{center}
\begin{tabular}{c c c}
\epsfig{file=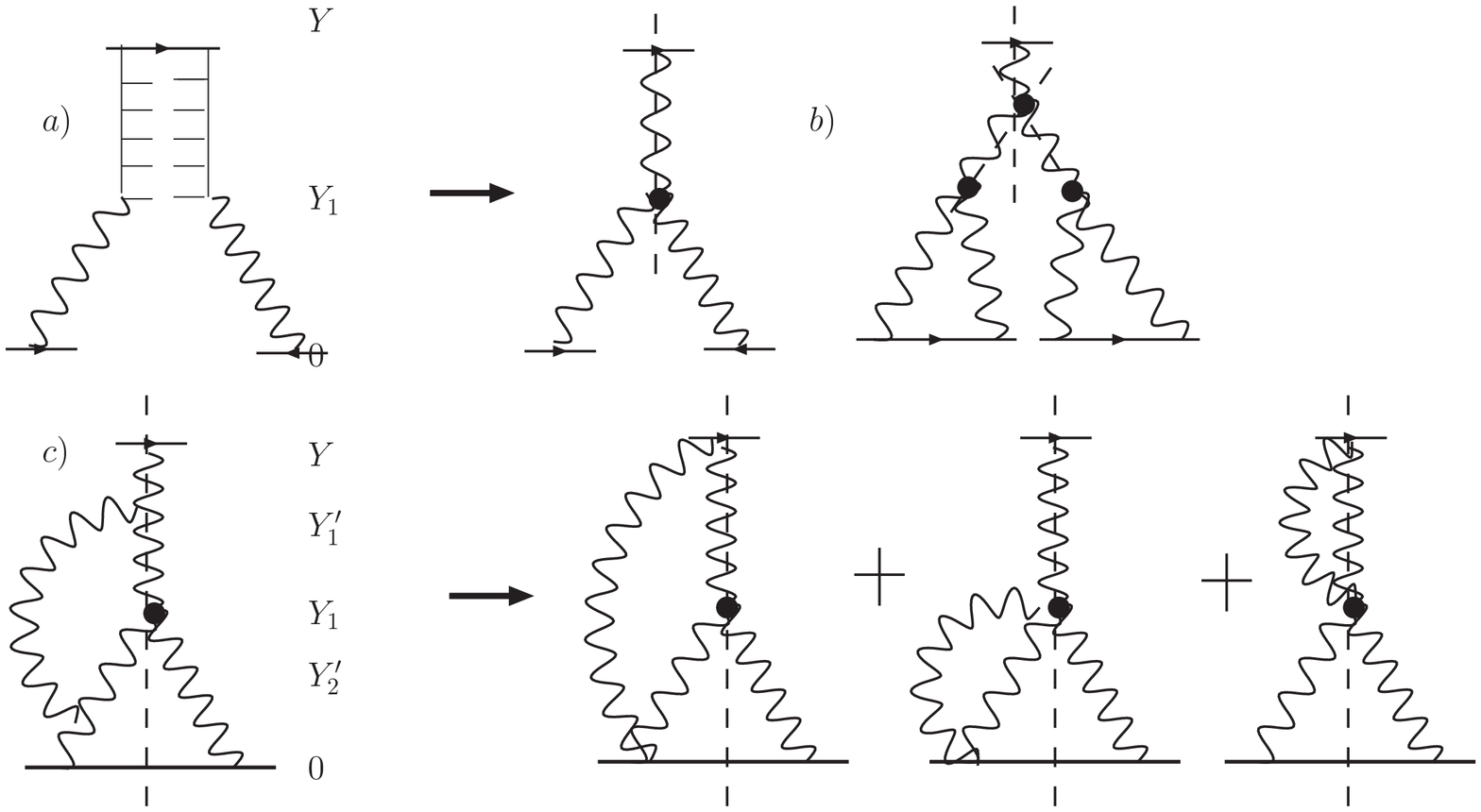,width=65mm,height=40mm}  
&\,\,\,\,\,\,\,\,&
\epsfig{file=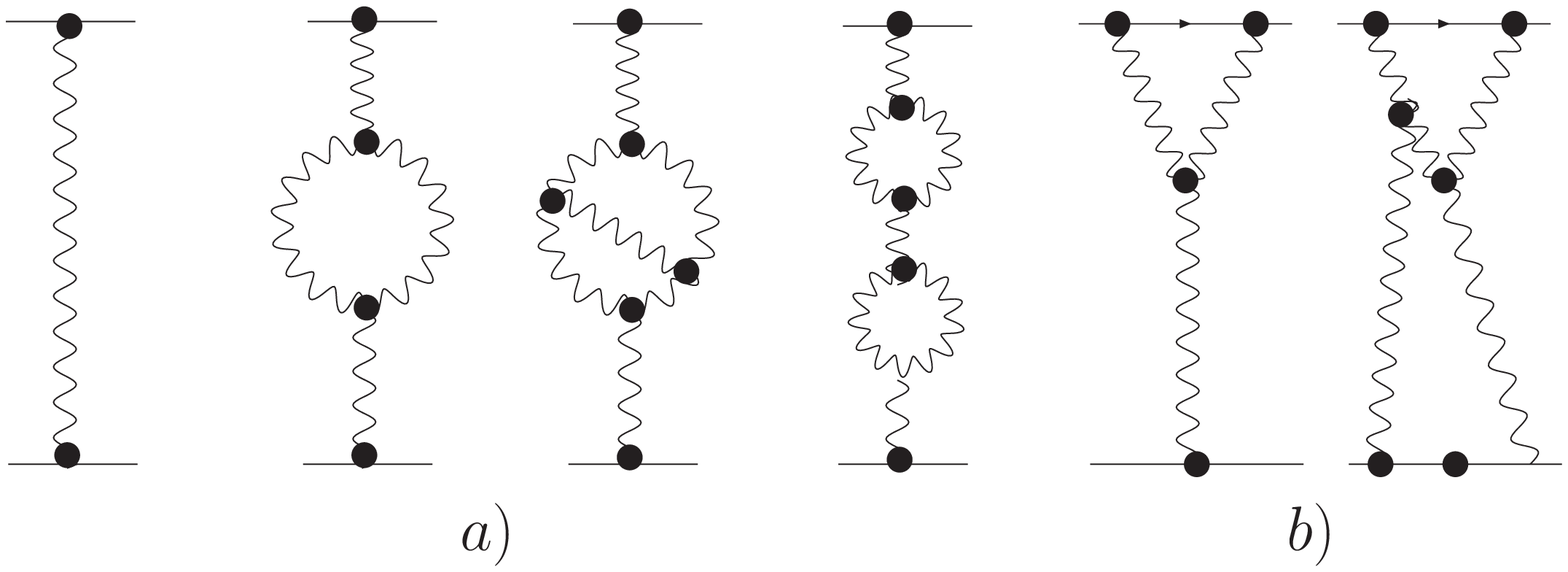,width=65mm,height=40mm} \\
\end{tabular}
\caption{L.H. figure: Examples of Pomeron diagrams not included in G-W 
mechanism.
R.H. figure: Low order terms of the Pomeron Green's
function. a) Enhanced. b) Semi-enhanced.}\label{pomdgrs}
\end{center}
\end{figure}

\par To simplify the problem of summing the Pomeron loop diagrams, we  
 assume that at high energies only the triple Pomeron interaction is 
essential, this conjecture has been proved in peturbative QCD (see 
Refs.[\cite{3P1} 
\cite{3P2}]). Mueller \cite{Muell2} has shown that in the leading log 
$x$ approximation of pQCD for a large number of colours ($N_{c}\gg$ 1), 
the correct degrees of freedom are colourless dipoles. 
\subsection{Summing Interacting Pomeron Diagrams}
  In the leading order approximation of pQCD, only one 
Pomeron (dipole) splitting into two Pomerons (dipoles), and two Pomerons
(dipoles) merging into one Pomeron (dipole) are considered.
 All other Pomeron vertices do not appear 
in the leading log $x$  approximation of pQCD. We therefore restrict 
ourselves to sum only Pomeron diagrams containing triple Pomeron vertices.
We add a caveat, that we neglect the 4$\pom$ term (which is needed for s 
channel unitarity), however, this term is only significant at energies 
W $ >   10^{5}$ , which is at the limit of the validity of our 
model.
\par To make the calculation tractable we further assume that the slope of 
the Pomeron trajectory $\alpha_{\pom}' = 0$. The results of our 
numerical fit to the relevant data, which we will discuss later,  
($\alpha_{\pom}'  = 0.01$) lends credence to this assumption.  

The theory which includes all the above ingredients can
be formulated in terms of a generating function\cite{MUCD,LELU}
\beq \label{Z}
Z(y,\,u)\,\,=\,\,\sum_n\,\,P_n(y)\,\,u^n,
\eeq
where, $P_n(y)$ is the probability to find $n$-Pomerons (dipoles) at
rapidity $y$.
At rapidity $y\,=\,Y=\ln(s/s_0)$  we can impose
an arbitrary initial condition. For example, demanding that
there is only one fastest parton (dipole), which is $P_1(y\,=\,Y)\,=\,1$,
while $P_{n>1}(y\,=\,Y)\,=\,0$. In this case we have the following
initial condition for the generating function
\beq \label{INC1}
Z(y\,=\,Y)\,=\,u\,.
\eeq
At $u =1$
\beq \label{INC2}
Z(y,\,u\,=\,1)\,\,=\,\,1,
\eeq
which follows from the physical meaning of $P_n$ as a probability.
The solution, with these two conditions, will give us the sum of enhanced
diagrams.

For the function $Z\Lb u\Rb $ the following simple equation can be written
(see Ref.\cite{GLMM} and references therein)
\beq \label{GFEQ}
\,\,-\frac{\partial\,Z(y,\,u)}{\partial\, \y}\,\,
=\,\,-\,\Gamma(1 \to 2)\,u\,(1\,-\,u)
\,\,\frac{\partial\,Z(y,\,u)}{\partial\, u}\,\,\,+\,\,\,
\Gamma(2 \to 1)\,u\,(1\,-\,u) \,\,
\frac{\partial^2\,Z(y,\,u)}{\partial^2\,u},
\eeq
where, $\Gamma(1 \to 2)$ describes the decay of one Pomeron (dipole)
into two Pomerons (dipoles), while
$\Gamma(2 \to 1)$ relates to the merging of
two Pomerons (dipoles) into one Pomeron (dipole).

Using the functional $Z$,
we  calculate the scattering amplitude\cite{LELU,K}, using the following 
formula:
\beq 
N\Lb Y\Rb\,\,\,\equiv\,\,\,\mbox{Im} A_{el}\Lb Y\Rb \,\,\,
=\,\,\,\sum^{\infty}_
{n =1}\frac{(-1)^n}{n!}\,\,\,
\frac{\partial^n\,Z(y,\,u)}{\partial^n\, u}|_{u =1}
\,\gamma_n(Y=Y_0,b), \label{Eqn:AMM}
\eeq
where, $\gamma_n(Y=Y_0,b)$ is the scattering amplitude of
$n$-partons (dipoles) at low energy.

The generating function approach
has the advantage that it can be solved analytically
(see Ref.\cite{KOLE}), using the MPSI \cite{MPSI} approximation.
The exact expression for the Pomeron Green's function is given by
\beq 
G_\pom \left(Y\right)\,\,\,=\,\,\,1\,\,-\,\,\exp \Lb \frac{1}{T(Y)}\Rb\,
\frac{1}{T(Y)}\,\,
\Gamma\Lb 0,\frac{1}{T(Y)} \Rb, \label{Eqn:AMMPSI}
\eeq
where $\Gamma \Lb 0,x \Rb$ is the incomplete gamma function, and
$
T\Lb Y\Rb\,\,\,=\,\,\gamma\,e^{\Delta_{\pom}\,Y}
$.
 $\gamma$ denotes the amplitude
of the two dipole interaction at low energy.
The MPSI approximation only takes into account the first term of the 
expression of the enhanced diagrams, neglecting other terms, as they are 
suppressed as $e^{-\Delta Y}$. Consequently, this approximation  is only 
reliable in the 
region $Y \leq min[\frac{1}{\gamma},\frac{1}{\alpha_{\pom}'m^{2}_{i}}]$.
\vspace{-0.3cm} 
\section{Determining the Parameters of the Model and Results of the Fit}
The pertinent details of our fit to the experimental data, and our
determination of the relevant parameters of the model, needed to describe
the soft interactions, are contained in \cite{GLMM}. In this section we
 only mention the salient features, and results of the fit.
\begin{table}[h]
\begin{tabular}{|l|l|l|l|l|l|l|l|l|}
\hline
& & &$\alpha_{\pom}^{\prime}$&
$g_1$ & $g_2$ &
$m_1$ & $m_2$ & \\
& $\Delta_\pom$ & $\beta$ &
$GeV^{-2}$ & $GeV^{-1}$ & $GeV^{-1}$ &
$GeV$ & $GeV$ & $\chi^2/d.o.f.$ \\
\hline
\,\,\,\,\,\,\,GW\,\,&0.120 & 0.46 & 0.012  &
1.27  & 3.33  & 0.913 & 0.98  &
\,\,\,\,\,\,0.87 \\
\hline
\,GW+$\pom-enh.$  & 0.335 & 0.34 &  0.010 & 5.82  &
239.6  & 1.54 & 3.06  &
\,\,\,\,\,\,1.00 \\
\hline
\end{tabular}
\caption{Fitted parameters for GLMM GW and GW+$\pom$-enhanced models.}
\label{t1}
\end{table}
\begin{figure}
\begin{center}
\includegraphics[width=65mm,height=50mm]{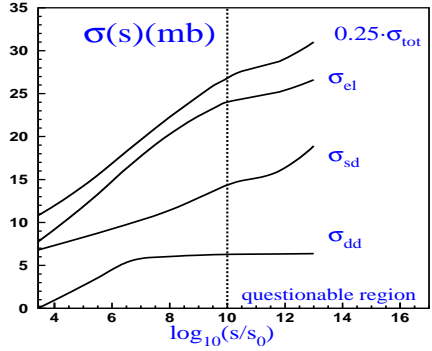}
\caption{ Energy dependence of GLMM cross sections.}\label{Energyxs}
\end{center}
\end{figure}
Our  fit is  based on 55 experimental data points,
which includes the $p$-$p$ and $\bar{p}$-$p$ total cross sections,
integrated elastic cross sections,
integrated single and double  diffraction cross sections,
and the forward slope of the elastic cross section
in the ISR-Tevatron energy range. The model gives
a good reproduction of the
data,  with a $\chi^2/d.o.f. \,\approx 1.$
In addition to the quantities contained in the data base,
we obtain a good description of the CDF \cite{CDF}
differential elastic cross sections and the single diffractive mass
distribution at $t\; =\; 0.05\;GeV^{2}$.  An important advantage
of our approach,  is that
the model provides a very good reproduction of the double diffractive (DD)
data points. Other attempts
to describe the DD data e.g.(see Refs.\cite{KMR08},\cite{KMR09}),
were not successful in reproducing the DD
experimental results over the whole energy range.

\par In Table 1  we list the values of the parameters obtained by a 
least squares fit 
to the experimental data, both for  the G-W formalism (elastic data 
only), 
and for the G-W formalism plus enhanced graphs (elastic plus diffractive 
data).  In Table 2 we compare our results
 with results of the
two versions of the model proposed by the Durham group \cite{KMR08} 
and\cite{KMR09}. In Fig.2 we display the predictions of the GLMM model's 
values for the various cross sections. Note that $\sigma_{el}$ and 
$\sigma_{sd}$ 
have completely different energy dependence, unlike the predictions of 
\cite{KMR09}. At an energy of 7 Tev the predictions of GLMM are (in mb): 
$\sigma_{tot}$= 86.0, $\sigma_{el}$= 19.5, $\sigma_{sd}$= 10.7,
$\sigma_{dd}$ = 5.9 and the forward slope $B_{el}$ = 19.4 $GeV^{-2}$.   
\begin{table}[ht]
\begin{tabular}{|l|l|l|l|}
\hline
& \,\,\,\,\,\,\,\,\,\,\,\,\,\,\,\,\,\,\,Tevatron
& \,\,\,\,\,\,\,\,\,\,\,\,\,\,\,\,\,\,\,\,\,\, LHC
& \,\,\,\,\,\,\,\,\,\,\,\, W=$10^5$ GeV  \\
& GLMM\,\,KMR(07)\,\,KMR(08)
& GLMM\,\,KMR(07)\,\,KMR(08)
& GLMM\,\,KMR(07)\,\,KMR(08) \\
\hline
$\sigma_{tot}$
&\,\, 73.3 \,\,\,\,\,\,\,\,\,\,74.0\,\,\,\,\,\,\,\,\,\,\,\,\,\,73.7
& \,\,\,\,92.1
\,\,\,\,\,\,\,\,\,\,\,88.0\,\,\,\,\,\,\,\,\,\,\,\,\,\,\,91.7
&\,108.0 \,\,\,\,\,\,\,\,\,\,\,98.0\,\,\,\,\,\,\,\,\,\,\,\,\,108.0 \\
\hline
$\sigma_{el}$
& \,\, 16.3 \,\,\,\,\,\,\,\,\,\,16.3\,\,\,\,\,\,\,\,\,\,\,\,\,\,16.4
& \,\,\,\,20.9
\,\,\,\,\,\,\,\,\,\,\,20.1\,\,\,\,\,\,\,\,\,\,\,\,\,\,\,21.5
& \,\,\,\,24.0 \,\,\,\,\,\,\,\,\,
22.9\,\,\,\,\,\,\,\,\,\,\,\,\,\,\,\,26.2 \\
\hline
$\sigma_{sd}$
& \,\,\,\,\, 9.8 \,\,\,\,\,\,\,\,\,\,10.9\,\,\,\,\,\,\,\,\,\,\,\,\,\,13.8
& \,\,\,\,11.8
\,\,\,\,\,\,\,\,\,\,\,13.3\,\,\,\,\,\,\,\,\,\,\,\,\,\,\,19.0
& \,\,\,\,14.4
\,\,\,\,\,\,\,\,\,\,\,15.7\,\,\,\,\,\,\,\,\,\,\,\,\,\,\,\,24.2 \\
\hline
$\sigma_{dd}$ & \,\,\,\,\, 5.4\,\,\,\,\,\,\,\,\,\,\,\,\,\,\,7.2
& \,\,\,\,\,\,\,6.1 \,\,\,\,\,\,\,\,\,\,\,13.4
& \,\,\,\,\,\,\,6.3\,\,\,\,\,\,\,\,\,\,\,\,\,17.3 \\
\hline
$\frac{\sigma_{el} + \sigma_{diff}}{\sigma_{tot}}$ & \,\,\,\, 
0.43\,\,\,\,\,\,\,\,\,\,\,\,\,\,0.46
& \,\,\,\,\,\,0.42 \,\,\,\,\,\,\,\,\,\,\,0.53
& \,\,\,\,\,\,0.41\,\,\,\,\,\,\,\,\,\,\,\,\,0.57 \\
\hline
\end{tabular}
\caption{Comparison of GLMM, KMR(07) and KMR(08) cross sections in $mb$.}
\label{T}
\end{table}
\vspace{-0.5cm}
\section {Survival Probability for Central Diffractive Production of the 
Higgs Boson}
      A general review of survival probability calculations can be found 
in \cite{GLMNP}.
    We denote by  $\langle \mid S^2 \mid\rangle$ the probability that the 
Large Rapidity Gap (LRG) survives, and is not filled by secondaries from 
eikonal and enhanced rescattering effects (see Fig. 3).
\begin{figure}
\begin{center}
\epsfig{file=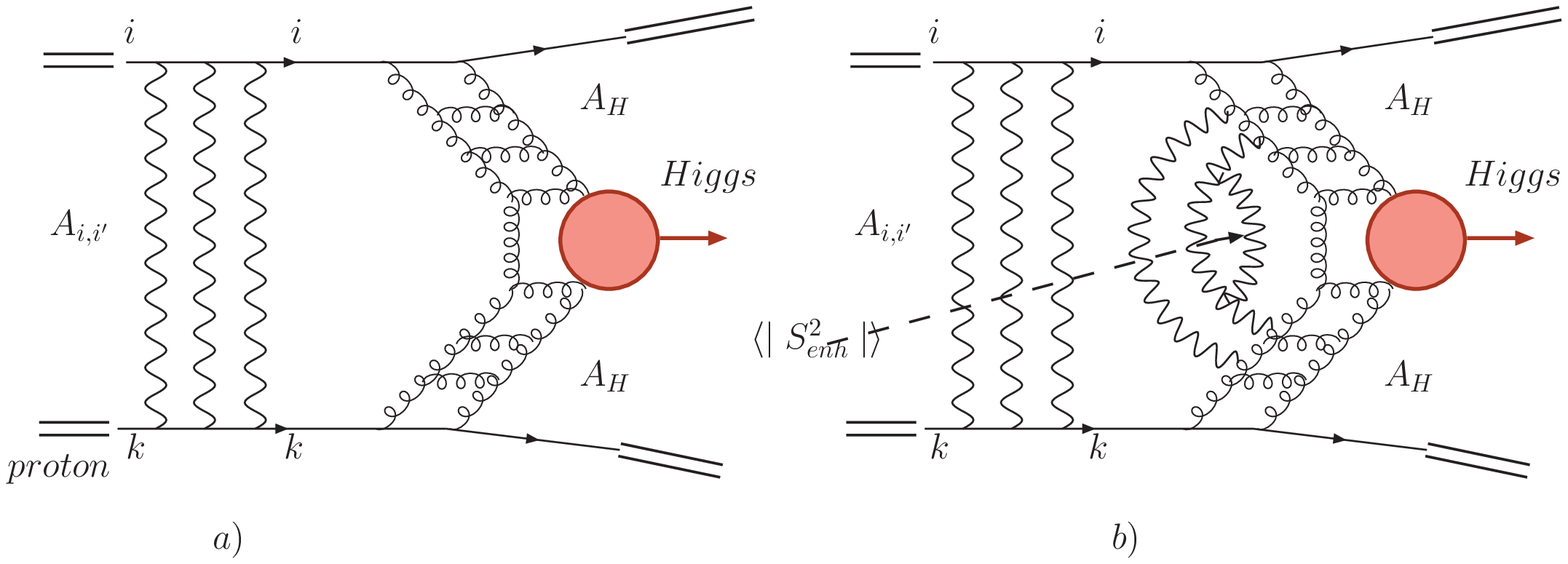,width=120mm,height=40mm}
\end{center}
\caption{ a) the survival probability in the G-W mechanism, 
b) illustrates the origin of the additional factor
 $\langle \mid S^2_{enh} \mid\rangle$.}
\label{sp1}
\end{figure}
 The expression for the survival probability can be written
$\langle\mid S^2_{2ch} \mid \rangle = \frac{N(s)}{D(s)}$
where

 $ N(s) = \int d^2\,b_1\,d^2\,b_2
\left[\sum_{i,k} \,<p|i>^2 <p|k>^2 \,A^{i}_H(s,b_1)\,A^k_H(s,b_2)
(1-A^{i,k}_S
\Lb(s,(\mathbf{b}_1+\mathbf{b}_2 )
\Rb)\right]^2,$

and

$D(s) = \int\,d^2\,b_1\,d^2\,b_2
\left[\sum_{i,k} <p|i>^2 <p|k>^2\, A^i_H(s,b_1)\,A^k_H(s,b_2)\right]^2$

$A_{S}(b,s)$ denotes the "soft" strong amplitude of our model \cite{GLMM}. 
While 
for the "hard" amplitude $A_{H}(b,s)$, we assume an input Gaussian b 
dependence. i.e.
${A_{i,k}^H} = A_H(s)\,\Gamma_{i,k}^H(b)$   $\;\;\;$
with $\;\;\;\;\;\;\;$
$\Gamma_{i,k}^H(b) =
\frac{1}{\pi (R^H_{i,k})^2}\,e^{-\frac{2\,b^2}{(R^H_{i,k})^2}}.$
The "hard" radii are constants determined from HERA data on elastic and
inelastic $J/\Psi$ production.
We introduce two hard b-profiles
$
A^{pp}_H(b)\; =\;
\frac{V_{p \to p}}{2 \pi B_{el}^H} \exp \Lb-\frac{b^2}{2\,B_{el}^H} \Rb,$
$\;\;$
 and $\;\;
 A^{pdif}_H(b)\; = \; \frac{V_{p \to dif}}{2 \pi
B_{in}^H} \exp \Lb -\frac{b^2}{2 B_{in}^H}\Rb.$
\begin{figure}
\begin{center}
\begin{tabular}{c c c}
\epsfig{file=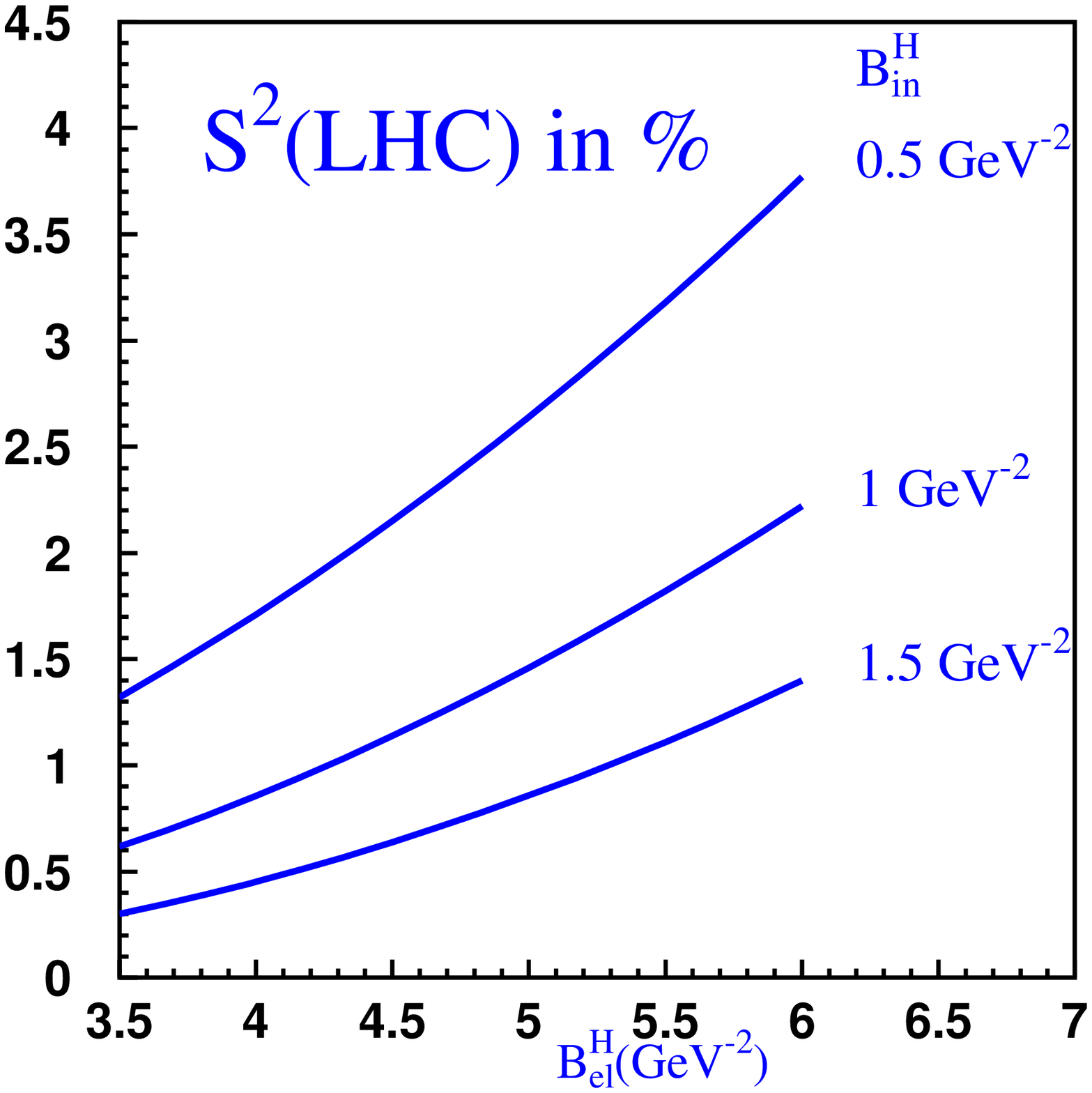,width=65mm,height=60mm}  
&\,\,\,\,\,\,\,\,&
\epsfig{file=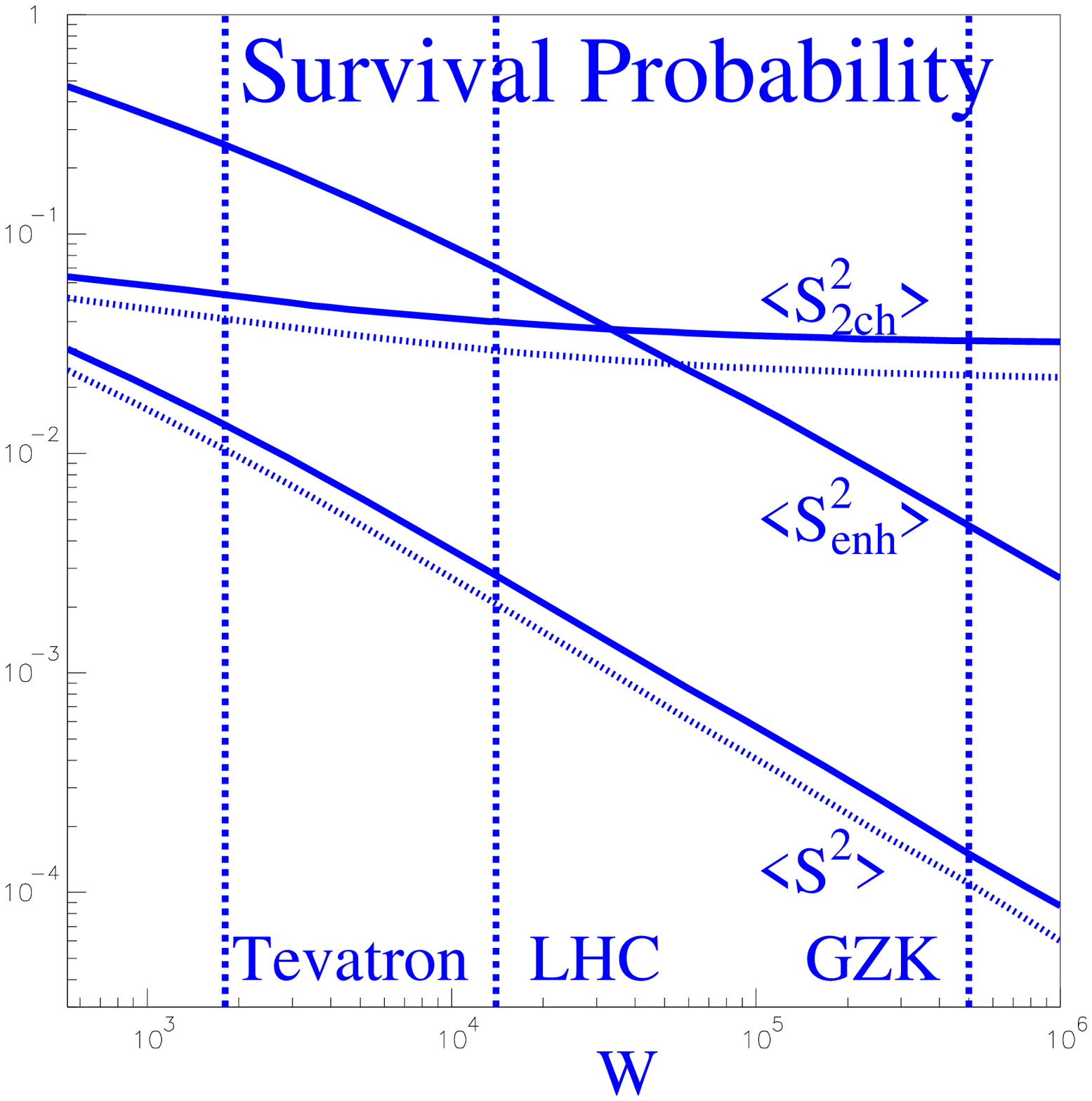,width=65mm,height=60mm} \\
\end{tabular}
\caption{L.H. figure: The dependence of $S^{2}$ at the LHC on $B^{H}_{el}$
and $B^{H}_{in}$.
R.H. figure: Energy dependence of $S^{2}$ for centrally produced Higgs.
The full (dashed) line is for $B^{H}_{el}$ = 5.0 (3.6) $GeV^{-2}$
.}\label{sp2}
\end{center}
\end{figure}
The values $B_{el}^H$=5.0 $GeV^{-2}$ and $B_{in}^H$=1 $GeV^{-2}$
have been taken from  ZEUS data. The value $B_{el}^H$= (3.6) $GeV^{-2}$
was used in \cite{GLMM}, this has now been changed in light of the latest 
measurements  of the "hard" slope, by the H1 group.
 This is in contrast to KMR treatment \cite{KMR08} where they assume:
$A^{pp}_{H}(b)\;=\; A^{pdif}_{H}(b)\;\propto\;$
$\exp \Lb -\frac{b^2}{2 B^H}\Rb$
with $B^{H}_{el}\; = \;B^{H}_{inel}$ = 4 or 5.5 $GeV^{-2}$
 The sensitivity of our results to the parameters of the "hard" 
amplitude are shown in Fig.4 (the L.H. figure), note that for
$B^{H}_{in}$ =1 $GeV^{-2}$, 
changing the value of $B^{H}_{el}$ from 3.6 to 5.0 $GeV^{-2}$, increases
$\langle\mid S^2_{2ch} \mid \rangle$ by $\approx$ 70 \% .
 Our results for
$\langle\mid S^2 \mid \rangle$ = $\langle\mid S^2_{2ch} \mid 
\rangle$ x $\langle\mid S^2_{enh} \mid \rangle$ is  given by the 
full line in Fig.4 (R.H. figure), it decreases with increasing energy, 
due to the behaviour of
$\langle\mid S^2_{enh} \mid \rangle$.  
 \par  Our results and the Durham group's results for Survival Probability
are given in Table 3.
\begin{table}[h]
\begin{tabular}{|l|l|l|l|}
\hline
& \,\,\,\,\,\,\,\,\,\,\,\,\,\,\,\,\,\,\,Tevatron
& \,\,\,\,\,\,\,\,\,\,\,\,\,\,\,\,\,\,\,\,\,\, LHC (14 TeV)
& \,\,\,\,\,\,\,\,\,\,\,\, W=$10^5$ GeV  \\
& GLMM\,\,KMR(07)\,\,KMR(08)
& GLMM\,\,KMR(07)\,\,KMR(08)
& GLMM\,\,KMR(07)\,\,KMR(08) \\
\hline
$S^2_{2ch}(\%)$ &
v\,\,\,\,\, 5.3 \,\,\,\,\,\,\,\,\,2.7-4.8 &
\,\,\,\,\,\,\,3.9  \,\,\,\,\,\,1.2-3.2 &
\,\,\,\,\,\,\,3.2  \,\,\,\,\,\,\,\,\,\,\,0.9-2.5 \\
\hline
$S^2_{enh}(\%)$ & \,\,\, 28.5 \,\,\,\,\,\,\,\,\,\,\,\,100
&\,\,\,\,\,\,\,6.3  \,\,\,\,\,\,\,\,\,\,\,\,100
\,\,\,\,\,\,\,\,\,\,\,\,\,\,\,33.3&
\,\,\,\,\,\,\,3.3 \,\,\,\,\,\,\,\, \,\,\,\,100 \\ \hline
$S^2(\%)$ &
\,\,\,\,\, 1.51  \,\,\,\,\,\,\,2.7-4.8
& \,\,\,\,\,\,\,0.24 \,\,\,\,\,\,\,1.2-3.2 \,\,\,\,\,\,\,\,\,\,\,\,1.5
&\,\,\,\,\,\,\,0.11\,\,\,\,\,\,\,0.9-2.5\\
\hline 
\end{tabular}
\caption{Comparison of results obtained for Survival Probability in Tel 
Aviv and Durham models}
\label{tab:t3}
\end{table}
 At an energy of 7 TeV we predict a value
$\langle\mid S^2 \mid \rangle$ $\approx$ 0.6\%. We have also succeeded in 
summing the semi-enhanced contribution (see R.H. side of Fig.1) to 
the Survival Probability, and find that it is almost energy independent, 
and has a value
$\approx$ 100\%  at Tevatron and LHC energies \cite{GLM5}.
\vspace{-0.5cm}
\section{Discussion and Conclusions}
We present a model for soft interactions having two components: 
(i) G-W mechanism for elastic and low mass diffractive scattering, and 
(ii) Pomeron enhanced contributions for high mass diffractive production.
In addition we find from our fit that the slope of the Pomeron
$\alpha_{\pom}'\approx
0.01$. This is consistent  with what one expects in pQCD, 
since for a BFKL Pomeron
$\alpha_{\pom}' \propto 1/Q^{2}_{s} \rightarrow \, 0$ as $s\,\,\rightarrow
\infty$.
Having $\alpha^{'}_{\pom}\,\,\rightarrow \,\,0$, provides a
necessary condition  that links
 strong (soft) interactions with the hard interactions described
by pQCD. A key hypothesis in our model is that the soft processes are not 
"soft", but originate from short distances.
 We have only one Pomeron. There is no requirement for a "soft" and "hard" 
Pomeron. This is in accord with Hera data for $F_{2}$, which is smooth 
throughout the transition region \cite{H1}.

\par To illustrate our achievements and problems,
we compare our approach with the work of the Durham group 
\cite{KMR08}\cite{KMR09}.
The main difference in the underlying philosophy of the two groups is
that,
the Durham approach is based on the parton model
where  there is only a short range rapidity interaction
between partons, while we, due to exchange of gluons in QCD,
have a long range rapidity interaction.
Both approaches consider $\alpha'_\pom$  as being small.
In both programs  the Pomeron interaction was taken into account.
The difference between the two approaches is that KMR
made an {\it ad hoc} "reasonable" assumption, that
the  multi-Pomeron vertices have the following form,
for the transition of  $n$ Pomerons to $m$ Pomerons
\beq \label{DG1}
g^n_m\,=\,n\,m\,\lambda^{n + m -2}\,g_N/2\,
=\,n\,m\,\lambda^{n+m-3}\,g_{3{\pom}}/2.
\eeq
No theoretical arguments or theoretical models were offered in support
of this assumption, which certainly contradicts
the pQCD approach \cite{3P1,3P2}.
In spite of these differences, the values obtained for the $\sigma_{tot}$ 
and $\sigma_{el}$ are in surprisingly close agreement (see Table 2). It is 
only in the latest version of the Durham model \cite{KMR09}, which 
includes three components of the Pomeron, with different transverse 
momenta of the partons in each component (to mimic BFKL diffusion 
in $k_{t}$), that there are fairly large discrepancies in the diffractive
sector i.e. $\sigma_{sd}$ and $\sigma_{dd}$. KMR  \cite{KMR09} find that 
at higher energies  $\sigma_{sd}$ and $\sigma_{el}$ have comparable values 
and similar energy dependence, this is not so in our description 
\cite{GLMM} (see Fig.2). We note that the results presented by Poghosyan
at this conference  for $\sigma_{sd}$ and $\sigma_{dd}$ \cite{POG}
 agree 
both in magnitude and energy dependence with those obtained in the GLMM
model.
 There is also disgreement in the result for the calculation of the 
Survival Probability. In \cite{KMR09}, the Survival Probability is now 
multiplied by a "renormalizing" factor
$(\langle p^{2}_{t} \rangle B)^{2}$ 
 and referred to as $\langle
S^{2}_{eff} \rangle$. The result for LHC energy is $\langle S^{2}_{eff} 
\rangle$
= $ 0.015^{+0.01}_{-0.005}$. It is not clear whether there is in addition 
a factor of  $\langle S^{2}_{enh}\rangle$
 $\approx 1/3$, that needs to be incorporated. If affirmative, then the 
discrepancy between our result and that of the Durham group for 
$\langle S^{2}\rangle$ at Tevatron energies is small, but the discrepancy 
becomes larger  for the LHC energy range, as we predict that $\langle 
S^{2}_{enh}\rangle$ decreases as the rapidity between the projectiles 
increases, while Durham claim little (if any) energy dependence for 
$\langle S^{2}_{enh}\rangle$.
\vspace{-0.5cm}
\section{Acknowledgments}
This research was supported in part by BSF grant $\#$ 20004019
and by a grant from Israel Ministry of Science
and the Foundation for Basic Research
of the Russian Federation.
\vspace{-0.5cm}
\section{Bibliography}
\begin{footnotesize}

\end{footnotesize}
\end{document}